\documentclass[pdflatex,sn-basic]{sn-jnl}
\usepackage{amsmath,amssymb,amsfonts,amsthm,mathrsfs}
\usepackage{graphicx,xcolor,textcomp}
\usepackage{bm,booktabs,multirow,array,enumitem}
\usepackage{algorithm,algpseudocode,float}
\usepackage[title]{appendix}
\usepackage{placeins}
\hypersetup{hidelinks,pdftitle={Asymptotic calibration of projected residual processes in multilevel linear mixed models},pdfauthor={Jakob Peterlin}}
\graphicspath{{figures/}}

\theoremstyle{thmstyleone}
\newtheorem{theorem}{Theorem}
\newtheorem{lemma}[theorem]{Lemma}
\newtheorem{proposition}[theorem]{Proposition}

\theoremstyle{thmstyletwo}

\newcommand{\R}{\mathbb{R}}

\DeclareMathOperator{\EE}{E}
\DeclareMathOperator{\COV}{cov}
\newcommand{\E}[1]{\EE\left[#1\right]}
\newcommand{\Cov}[1]{\COV\left(#1\right)}

\DeclareMathOperator{\col}{col}

\newcommand{\convP}{\overset{p}{\longrightarrow}}

\DeclareMathOperator{\rank}{rank}

\newcommand{\figalt}[1]{}

\input{labels-supplement.tex}

\begin{document}

\title[Projected residual processes in multilevel mixed models]{Asymptotic calibration of projected residual processes in multilevel linear mixed models}

\author*[1]{\fnm{Jakob} \sur{Peterlin}}\email{jakob.peterlin@mf.uni-lj.si}
\affil*[1]{\orgdiv{Institute for Biostatistics and Medical Informatics}, \orgname{Faculty of Medicine, University of Ljubljana}, \orgaddress{\street{Vrazov trg 2}, \city{Ljubljana}, \postcode{1000}, \country{Slovenia}}}

\abstract{A cumulative residual test checks a fitted regression model: residuals are added up in the order of a fitted prediction, and their path is compared with paths simulated under the model. A path that stands out indicates lack of fit and shows where it occurs. For linear mixed models the large-sample theory covers one grouping factor and simulates by perturbing independent groups. With nested or crossed factors such groups may be few or absent. We therefore project residuals off all fixed and random design columns, and order the observations by the fixed prediction alone or with predicted random effects added. The ordering determines what a test can detect and what its calibration requires. Along the fixed prediction the test is sensitive to the mean; for smooth weights and under conditions on design and fit, it is valid for independent errors, Gaussian or not, with simulated paths from fresh draws of any standardised law. In balanced designs with complete repeated profiles, one multiplier per profile keeps it valid for any covariance within a profile. Orderings that include random effects can reveal an omitted random slope; there we assume Gaussian errors, and Gaussian draws are valid in any design as the residual dimension grows. In simulations with nested and crossed factors, the tests rejected 4.2--5.7\% of null samples at the 5\% level; an omitted quadratic term was detected best along the fixed prediction and an omitted subject slope along the subject ordering. Crossed lexical-decision data illustrate the tests.}



\maketitle

\section{Introduction}\label{sec:intro}

In a longitudinal study, an unexplained bend in the mean trajectory and
an omitted subject-specific slope call for different changes to a mixed
model. Either can leave a pattern in a plot of cumulative residuals. Such
a plot shows a \emph{cumulative residual process}: the observations are
ordered by a fitted prediction, their residuals are added up in that
order, and the resulting path is compared with simulated paths
\citep{Su91,stute97,Lin02}. We call the process computed from the data
the \emph{primary process} and the simulated ones \emph{secondary
processes}. What a pattern means depends on which residuals are
accumulated, how the observations are ordered and how the secondary
processes are generated.

\citet{Pan05} use cumulative residuals to assess generalised linear mixed
models along covariates or fitted values. For linear mixed models,
\citet{TangSludPfeiffer2014} compare observed responses with the fitted
mean over a partition of the covariate space. \citet{LMMpaper} developed
cumulative residual tests for linear mixed models with one grouping
factor; their secondary processes multiply the residuals of each
independent group by a random multiplier.

With several grouping factors, independent groups may be few or absent.
Throughout, \emph{subjects} are the lowest grouping factor. In a nested
design they lie within \emph{centres}, and there may be only a few
centres. In a crossed design every subject meets several \emph{items},
and the same items link different subjects. We therefore project the
residuals off all fitted fixed and random design columns together. This
gives one set of residuals and a family of processes $O_S$ that differ in
their ordering. The process $O_\varnothing$ orders by the fixed
prediction; $O_S$ adds the predicted random effects of the factors in a
nonempty set $S$ and can show a departure along a particular random
effect. The question is then which arguments for the secondary processes
need the grouping structure and which need only the projection.

\begin{figure}[ht]
\centering
\includegraphics[width=.9\linewidth]{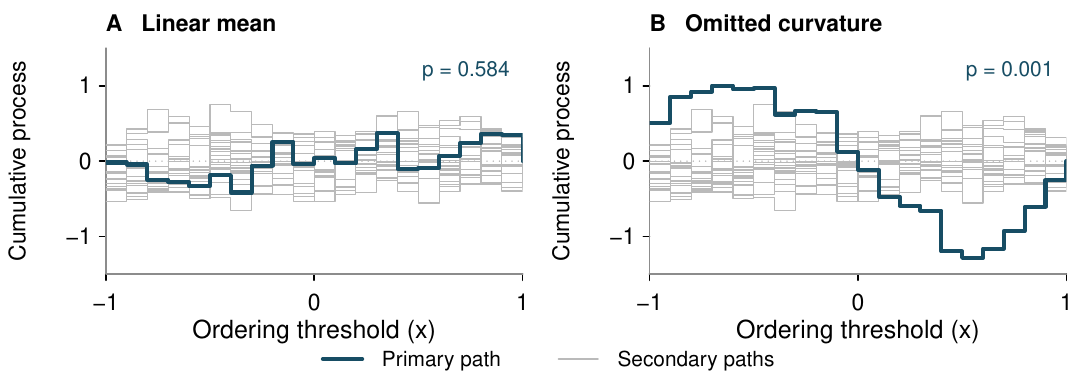}
\caption{The test for $O_\varnothing$. Three subjects with offsets
$-0.8$, $0$ and $0.8$ share 21 equally spaced covariate values $x$; the
mean is linear in $x$ in A and adds $1.4x^2$ in B. Both panels use the
same Gaussian errors ($\sigma=0.5$) and order by the fitted line,
equivalently by $x$. The dark step path is the primary path; the 19 grey
paths are secondary paths from independent random signs. Each root CvM $p$-value uses all 999
secondary paths with indicator weights. The fitted slope is positive in
both samples, so all indicator cuts agree with those of $x$
(Section~\ref{sec:balanced}).}
\label{fig:intro-projection}
\end{figure}

Figure~\ref{fig:intro-projection} shows the test for $O_\varnothing$.
Giving each subject its own intercept makes the projection remove the
subject offsets, but not a curvature that the subjects share. With a linear mean, the
primary path fluctuates among the secondary paths. Adding curvature to the
same observations produces two sustained excursions. The root Cram\'er--von
Mises (CvM) statistic measures the size of a whole path by its root mean
square over the fitted ordering values.

Our results are organised by the ordering; Section~\ref{sec:model}
defines the model, the processes and the tests.
\begin{enumerate}[label=(\roman*),nosep]
\item \emph{Fixed-prediction ordering} (Section~\ref{sec:fixed-order}).
The primary process $O_\varnothing$ has a Gaussian limit under independent
errors with a common variance and bounded fourth moments, Gaussian or
not, and secondary processes from fresh independent draws of any centred,
unit-variance law share it (Theorem~\ref{thm:fixed-path}).
\item \emph{Unit multipliers} (Section~\ref{sec:unit-reference}).
Multiplying the observed contribution of a whole unit by one random draw
keeps the dependence within the unit. In balanced designs with complete
repeated profiles, one multiplier per profile is valid for any covariance
within a profile, even when the working covariance is wrong
(Theorem~\ref{thm:whole-profile}).
\item \emph{Random-effect orderings} (Section~\ref{sec:theory}). Here the
ordering uses the same errors as the residuals, and we assume Gaussian
errors. Oracle, primary and secondary processes then share one limit
(Theorem~\ref{thm:gauss-path}), and Gaussian draws are valid in any
design as the residual dimension grows
(Proposition~\ref{prop:gaussian-O-main}): they differ only by a radial
factor from the exact rotation test (Section~\ref{sec:exact-link}).
\item \emph{Detection} (Section~\ref{sec:detection}). Mean functions
describe which departures each ordering can show. A covariance error
outside the fitted span can inflate the variance of the fixed-prediction
process while its mean path is zero.
\end{enumerate}
Section~\ref{sec:simulations} reports simulations and an analysis of
crossed lexical-decision data.

\section{Model, processes and tests}
\label{sec:model}

\subsection{The model}
The design $\mathcal D_n$ records the covariates and group memberships;
we treat it as given. The working Gaussian likelihood for maximum
likelihood (ML) or restricted maximum likelihood (REML) uses
\begin{equation}
 \bm y=\bm X\bm\beta+\sum_{l=1}^L\bm Z_l\bm b_l+\bm\epsilon,
 \qquad \bm b_l\sim N(\bm0,\sigma^2\bm\Lambda_l),\qquad
 \bm\epsilon\sim N(\bm0,\sigma^2\bm I),
 \label{eq:model-main}
\end{equation}
with independent terms, $\sigma>0$ and $\bm\Lambda_l=\bm I_{m_l}\otimes\bm
D_l$, $\bm D_l\succeq0$. Here $\bm X\in\R^{n\times p}$ has full column rank;
grouping factor $l$ has $m_l$ groups and $k_l$ random coefficients per
group, with covariance $\sigma^2\bm D_l$, and $\bm Z_l\in\R^{n\times m_lk_l}$
is its design. The numbers $L$, $p$ and $k_1,\ldots,k_L$ are fixed, the
group counts may grow with $n$ or stay fixed, and the factors may be nested
or crossed in any way. Let $\bm Z=[\bm Z_1,\ldots,\bm Z_L]$ and, for a set
$S$ of factors ($\bm C_\varnothing=\bm0$),
\begin{equation}
 \bm R=\bm I-[\bm X,\bm Z][\bm X,\bm Z]^+,\quad
 \bm C_S=\sum_{l\in S}\bm Z_l\bm\Lambda_l\bm Z_l^{\mathsf T},\quad
 \bm V=\bm I+\bm C_{\{1,\ldots,L\}},\quad
 \bm H_S=\bm C_S\bm V^{-1},
 \label{eq:matrices-main}
\end{equation}
where $^+$ denotes the Moore--Penrose inverse. We call
$\col[\bm X,\bm Z]$, the span of the fixed and random-effect design
columns, the \emph{fitted span}. Thus $\sigma^2\bm V$ is the covariance of
$\bm y$ and $\bm R$ the orthogonal projection onto the orthogonal
complement of the fitted span, so that $\bm R\bm X=\bm0$,
$\bm R\bm Z=\bm0$ and $\bm R\bm H_S=\bm0$. With $\bm e=\bm y-\bm X\bm\beta$,
the vector $\bm H_S\bm e$ is the contribution of the factors in $S$ to the
best linear unbiased prediction (BLUP) \citep{Robinson91}, and
$\bm R\bm e=\bm R\bm\epsilon$ contains neither fixed nor random effects. We
assume $r=\rank\bm R>0$. Model \eqref{eq:model-main} is the working model
of the fit. Section~\ref{sec:fixed-order} needs less: independent errors
that need not be Gaussian, and a random component that lies in the fitted
span.

\subsection{The projected processes}
A process accumulates \emph{marks}, one per observation, in the order given
by \emph{ordering values}. For a set $S$ of factors the process $O_S$
orders by the fixed prediction plus the predicted contribution of the
factors in $S$, and its marks are the residuals projected by $\bm R$. At
the true parameters,
\begin{equation}
 \bm y^{O_S}=\bm X\bm\beta+\bm H_S\bm e,\qquad
 \bm e^{O}=\sigma^{-1}\bm R\bm e,
 \label{eq:orderings-main}
\end{equation}
and the \emph{oracle process}, a theoretical benchmark, is
\begin{equation}
 W_n^a(t)=n^{-1/2}\bm\sigma_\lambda(\bm y^a,t)^{\mathsf T}\bm e^{O},
 \qquad t\in\R,\quad a=O_S .
 \label{eq:processes-main}
\end{equation}
Here $\lambda\in(0,\infty]$ is the smoothing \emph{scale},
$\sigma_\lambda(x)=1$ for $x\le0$,
$1-10(\lambda x)^3+15(\lambda x)^4-6(\lambda x)^5$ for $0<x<1/\lambda$ and
$0$ for $x\ge1/\lambda$, with $\sigma_\infty(x)=1\{x\le0\}$, is the quintic
weight of \citet{LMMpaper}, and $\bm\sigma_\lambda(\bm v,t)$ is the vector
with coordinates $\sigma_\lambda(v_i-t)$. With indicator weights
($\lambda=\infty$) the process is $n^{-1/2}$ times the sum of the marks of
the observations ordered at or below the threshold $t$; a finite scale
lets each mark enter gradually over a window of width $1/\lambda$. The
\emph{primary process}, computed from the data as in \citet{LMMpaper}, is
\begin{equation}
 \widehat W_n^a(t)=n^{-1/2}\bm\sigma_\lambda(\widehat{\bm y}^a,t)^{\mathsf T}
 \widehat{\bm e},\qquad
 \widehat{\bm y}^{O_S}=\bm X\widehat{\bm\beta}+\sum_{l\in S}\bm Z_l\widehat{\bm b}_l,
 \qquad
 \widehat{\bm e}=\frac{\bm R\bm y}{\widetilde\sigma},
 \label{eq:fitted-process-main}
\end{equation}
where $\widetilde\sigma=\|\bm R\bm y\|/\sqrt r$, hats denote ML estimates,
or REML covariance estimates with their generalised least squares (GLS)
fixed effects, and
$\widehat{\bm b}_l=\widehat{\bm\Lambda}_l\bm Z_l^{\mathsf T}
\widehat{\bm V}^{-1}(\bm y-\bm X\widehat{\bm\beta})$ are the predicted random
effects of factor $l$. A \emph{path} is a process as a function of $t$.

All processes share the \emph{primary marks} $\widehat{\bm e}$ and differ
only in the ordering. We call $\bm X\widehat{\bm\beta}$ the
\emph{fixed-prediction ordering} and $O_\varnothing$ the fixed-prediction
process; an ordering with $S\ne\varnothing$ is a \emph{random-effect
ordering}. We write $O=O_{\{1,\ldots,L\}}$ for all factors together and
$O_l=O_{\{l\}}$ for factor $l$ alone. With subjects crossed with items,
for example, $O_{\rm subj}$ orders by the fixed prediction plus the
predicted subject effects.

Projecting off $\bm Z$ alone, as in the one-factor construction, would
leave $(\bm I-\bm Z\bm Z^+)\bm X(\bm\beta-\widehat{\bm\beta})$ in the fitted
residual. Including $\bm X$ in the projection removes that estimation term
at every sample size. Since $\bm R\bm X=\bm0$ and $\bm R\bm Z=\bm0$, the
primary marks are $\bm R\bm\epsilon/\widetilde\sigma$: they depend on no
fitted parameter, and their squared length is exactly $r$. The square of
the \emph{residual standard deviation} $\widetilde\sigma$,
$\widetilde\sigma^2=\bm y^{\mathsf T}\bm R\bm y/r$, is the residual mean
square of the least-squares fit of $\bm y$ on $[\bm X,\bm Z]$, which is
unbiased for $\sigma^2$ whatever the random effects. With random intercepts
only, the mark of an observation is roughly its residual centred within its
groups.

A process is \emph{degenerate} when its weight vectors
$\bm\sigma_\lambda(\widehat{\bm y}^a,t)$ lie in the fitted span, so that
its paths vanish identically. This happens when the ordering is constant on
sets of observations whose indicators lie in that span: the groups of a
factor with a random intercept, as with group-level covariates only, or the
observations of one group at one value of a binary covariate that has a
random slope. With $\bm L_a$ the indicator matrix of the blocks of tied
ordering values, the process is degenerate at every scale when
$\mathrm{df}_a=\rank(\bm R\bm L_a)=0$. The computed paths are then rounding
errors that still give a Monte Carlo $p$-value, and an untestable model is
not thereby confirmed; Algorithm~\ref{alg:test} checks this.

\subsection{Secondary processes}
\label{sec:secondary}
A secondary process keeps the fitted ordering and scale and replaces the
primary marks by simulated ones. We use three constructions
(Table~\ref{tab:constructions}). For \emph{fresh draws}, generate for each
draw $b=1,\ldots,B$ an independent vector $\bm\xi_b$ whose coordinates are
independent and identically distributed (iid) with mean zero and variance
one, and put
\begin{equation}
 \widehat W_{n,b}^{a\#}(t)=n^{-1/2}
 \bm\sigma_\lambda(\widehat{\bm y}^a,t)^{\mathsf T}\bm R\bm\xi_b .
 \label{eq:secondary-main}
\end{equation}
Standard normal coordinates give \emph{Gaussian draws}; random signs are
another choice. A \emph{rotation} is a Gaussian draw normalised to the
length of the primary marks,
\begin{equation}
 \sqrt r\,\bm R\bm\xi_b/\|\bm R\bm\xi_b\|,\qquad
 \bm\xi_b\sim N(\bm0,\bm I),
 \label{eq:exact-draw-main}
\end{equation}
used as the marks in \eqref{eq:secondary-main}. Under Gaussian errors it
reproduces the conditional law of the primary marks exactly. The exact
test that it gives is the subject of the companion paper
\citep{exactpaper}; here it is the benchmark for Gaussian draws
(Section~\ref{sec:exact-link}). \emph{Unit multipliers}
(Section~\ref{sec:unit-reference}) do not generate new marks: they
multiply the observed contribution of each unit of observations by one
random draw.

\begin{table}[tb]
\caption{The three constructions of secondary processes and where each
is proved valid; the multipliers $\xi_{b,U}$ of the units $U$ are iid with
mean zero and variance one.}
\label{tab:constructions}
\centering\footnotesize
\renewcommand{\arraystretch}{1.1}
\begin{tabular}{@{}l>{\raggedright\arraybackslash}p{.27\textwidth}>{\raggedright\arraybackslash}p{.47\textwidth}@{}}
\toprule
Construction & Secondary marks & Proved valid for\\
\midrule
Fresh draws & $\bm R\bm\xi_b$
 & $O_\varnothing$ with smooth weights under independent errors
 (Theorem~\ref{thm:fixed-path}); Gaussian draws for every $O_S$ under
 Gaussian errors (Theorem~\ref{thm:gauss-path},
 Proposition~\ref{prop:gaussian-O-main})\\
Rotations & $\sqrt r\,\bm R\bm\xi_b/\|\bm R\bm\xi_b\|$
 & every $O_S$ under Gaussian errors, exactly at every sample size; the
 benchmark for Gaussian draws\\
Unit multipliers & $\bm R\bm g_b/\widetilde\sigma$,
 $(\bm g_b)_U=\xi_{b,U}(\bm R\bm y)_U$
 & $O_\varnothing$ with complete repeated profiles, any covariance within
 a profile (Theorem~\ref{thm:whole-profile}); with a leverage correction,
 every $O_S$ under Gaussian errors
 (Proposition~\ref{prop:gauss-unit})\\
\bottomrule
\end{tabular}
\end{table}

\subsection{Statistics, $p$-values and the minimum-$p$ rule}
\label{sec:pvalues}
The Kolmogorov--Smirnov (KS) and root CvM statistics evaluate a path at
the fitted ordering values, which we call the \emph{fitted grid},
\begin{equation}
 T_{{\rm KS},n}(W)=\max_i|W(\widehat y_i^a)|,\qquad
 T_{{\rm CvM},n}(W)=\Bigl\{n^{-1}\textstyle\sum_iW(\widehat y_i^a)^2\Bigr\}^{1/2}.
 \label{eq:statistics-main}
\end{equation}
KS records the largest absolute excursion. Observations with equal
ordering values are kept together.
With the statistic $T_0$ of the primary process and $T_1,\ldots,T_B$ of the
secondary processes, the Monte Carlo $p$-value is
\begin{equation}
 \widehat p=\Bigl\{1+\textstyle\sum_{b=1}^B1\{T_b\ge T_0\}\Bigr\}/(B+1).
 \label{eq:mc-main}
\end{equation}
Small $\lambda$ emphasises broad patterns, large $\lambda$ local variation.
For each process let $I_n$ be the range between the 5th and 95th
percentiles of its fitted ordering values and use the relative scales
$\lambda=c/I_n$ with $c\in\mathcal C=\{1,2,10,100,1000\}$, so that $c=1$
smooths over most of the range and $c=1000$ is close to indicator weights,
which are reported alongside.

Which scale shows a departure is not known in advance. We take the smallest
$p$-value over the scales and calibrate that minimum with the same draws,
the single-step minimum-$p$ adjustment of \citet{WestfallYoung1993};
\citet{LMMpaper} maximised standardised statistics instead. A
\emph{column} is a combination of a process, a scale and a statistic, and
a \emph{row} collects the statistics of one process over the columns:
row $b=0$ is the \emph{primary row} and rows
$b=1,\ldots,B$ are the secondary rows. With $T_{bc}$ the statistic of
row $b$ in column $c$,
\begin{equation}
 p_{bc}=\frac{\sum_{b'=0}^B1\{T_{b'c}\ge T_{bc}\}}{B+1},\qquad
 M_b=\min_{c\in\mathcal C}p_{bc},\qquad
 \widehat p_{\min}=\frac{\sum_{b=0}^B1\{M_b\le M_0\}}{B+1}.
 \label{eq:minp-main}
\end{equation}
Each secondary row repeats the selection made for the primary row. The
columns may be the five scales of one process, or any prespecified set of
processes, scales and statistics; one draw then serves every column of
its row, and the construction must reproduce the joint law across the
columns. The following bound is the standard validity result for Monte
Carlo tests \citep{Barnard1963,Hemerik2018}, with a lower bound for the
ties that the minimisation creates; exchangeability of the rows means that
permuting the row labels leaves their joint law unchanged.

\begin{proposition}[Rank bounds for the minimum-$p$ rule]
\label{prop:finite-rank-main}
Let $N=B+1$, let $\mathcal C$ be any fixed set of $J\ge1$ columns and
$k_\alpha=\lfloor\alpha N\rfloor$ for
$0<\alpha<1$. If the $N$ rows $(T_{bc})_{c\in\mathcal C}$, $b=0,\ldots,B$,
are exchangeable, then $\Pr(\widehat p_{\min}\le\alpha)\le k_\alpha/N$. If,
in addition, the $N$ statistics of each column are pairwise distinct almost
surely, then
$\max\{0,(k_\alpha+1-J)/N\}\le\Pr(\widehat p_{\min}\le\alpha)\le k_\alpha/N$.
No independence between columns is required.
\end{proposition}

For $B=999$ and five scales the bounds are $[0.046,0.050]$ at
$\alpha=0.05$ when the columns have no ties. The same bounds hold in the
limit if the primary row and the conditional law of a secondary row
converge to the same deterministic law with continuous marginals. For
$B_n\to\infty$ the minimum-$p$ value then tends in law to uniformity
(Supplementary Proposition~\ref{S-prop:joint-statistic-ranks}).

\paragraph{Choosing the construction.}
Table~\ref{tab:constructions} and step~3 of Algorithm~\ref{alg:test} give
the choice. Under Gaussian errors we use Gaussian draws, valid for every
ordering once the residual rank $r$ is large; the rotation test is
exact at every $r$ and preferable when $r$ is small. Without Gaussian errors, fresh draws serve the fixed-prediction
ordering under independent errors, and profile multipliers when the
covariance within profiles is unknown. A joint test needs one
construction that is justified for all its columns. Our main test is
root CvM with the minimum-$p$ rule over the five relative scales. The
projection requires one factorisation of $[\bm X,\bm Z]$.

\begin{algorithm}[tb]
\caption{The test on the fitted grid}\label{alg:test}
\begin{algorithmic}[1]
\State Fit the model. Compute $\bm R\bm y$, $\widetilde\sigma$, the chosen
orderings and their scales. If $\bm R\bm y=\bm0$ there is no test.
\State Drop orderings with $\rank(\bm R\bm L_a)=0$. Compute the primary
row from \eqref{eq:fitted-process-main}.
\State Generate $B$ secondary rows: Gaussian draws
\eqref{eq:secondary-main} under Gaussian errors, or their rotations
\eqref{eq:exact-draw-main} when $r$ is small; fresh draws for smooth
fixed-prediction columns under independent errors
(Theorem~\ref{thm:fixed-path}); or profile multipliers \eqref{eq:scores}
(Theorem~\ref{thm:whole-profile}). Share each draw across columns.
\State Rank the primary row by \eqref{eq:mc-main} or by the minimum-$p$
rule \eqref{eq:minp-main}. Plot the primary path with secondary paths.
\end{algorithmic}
\end{algorithm}

\section{The fixed-prediction process under independent errors}
\label{sec:fixed-order}

For the fixed-prediction ordering write
$x_i^{\mathsf T}\beta=\beta_0+\tilde x_i^{\mathsf T}\gamma$, where
$\tilde x_i$ is the design row without its intercept entry and $\gamma$
collects the slopes, and measure thresholds from the intercept: $t$ below
corresponds to $\beta_0+t$ for the oracle process and to
$\widehat\beta_0+t$ for the primary process. Subtracting the intercept from
the predictions and from the thresholds changes no weight on the fitted
grid. It therefore changes neither statistic, no rank and not the central
range used for the relative scales, and the intercept needs no rate of
convergence. If the model has no intercept, use the ordering as fitted. Set
$w_{\lambda,\gamma,i}(t)=\sigma_\lambda(\tilde x_i^{\mathsf T}\gamma-t)$ and
$c_{\lambda,\gamma}(t)=\bm Rw_{\lambda,\gamma}(t)$. For a fixed design,
the oracle process for $O_\varnothing$ has the representation
\begin{equation}
 W_{n,\lambda}(t)=
 \frac{1}{\sqrt n}\sum_i c_{\lambda,\gamma,i}(t)\,\epsilon_i/\sigma,
 \qquad
 K_{jk,n}(s,t)=n^{-1}c_{\lambda_j,\gamma}(s)^{\mathsf T}
                         c_{\lambda_k,\gamma}(t).
 \label{eq:fixed-kernel}
\end{equation}
This follows from $\bm R\bm y=\bm R\bm\epsilon$ and symmetry of $\bm R$.
The errors in the sum are independent even if every pair of subjects shares
an item. Thus the mean of the oracle process is zero and its covariance is
$K_{jk,n}$ for any independent centred errors with common variance
$\sigma^2$. The primary process uses estimated weights and the residual
standard deviation $\widetilde\sigma$.

\begin{theorem}[Fixed-prediction paths and fresh draws]
\label{thm:fixed-path}
Condition on a sequence of designs. Suppose the observation errors are
independent, centred, independent of the random effects, with variance
$\sigma^2>0$ and uniformly bounded fourth moments. Suppose $r\to\infty$,
$p$ is fixed, the design rows $\bm x_i$ are uniformly bounded and
$\|\widehat\gamma-\gamma\|=o_p(n^{-1/6})$.
Take a fixed finite collection of $J$ smooth scales, each either a positive
constant or $c/I_n$ with $c>0$ and $I_n$ the central 90\% range of the
fitted fixed predictions. For an oracle relative scale use $c/I_n^0$,
where $I_n^0$ is the corresponding range of the oracle predictions.
Assume $I_n^0$ is bounded away from zero. On a fixed compact interval $T$
that contains the oracle predictions and, with probability tending to one,
the fitted predictions, both measured from their intercepts, suppose
\begin{equation}
 \max_{j,i}\sup_{t\in T}|c_{\lambda_j,\gamma,i}(t)|/\sqrt n\longrightarrow0,
 \qquad K_{jk,n}(s,t)\longrightarrow K_{jk}(s,t)
 \quad(s,t\in T).
 \label{eq:fixed-diffuse}
\end{equation}
Then the oracle and primary paths, measured from their intercepts, converge
jointly over the scales in $C(T)^J$ to the same centred Gaussian process
with kernel $K$, and their supremum-norm difference tends to zero in
probability. Given the data, the secondary paths \eqref{eq:secondary-main}
from fresh draws converge in law to this process, in probability, for any
iid law of the draws with mean zero and variance one, independent of the
data.
\end{theorem}

In particular Gaussian draws, random signs and standardised asymmetric
two-point draws all give the same limit. The first condition in
\eqref{eq:fixed-diffuse} asks that no single error carries a nonvanishing
share of a path; we then call the projected coefficients \emph{diffuse}. A
bounded maximum absolute row sum of $\bm R$ suffices. The assertion is
about the distribution under repeated samples at the given design.
Conditioning additionally on the fitted ordering of a non-Gaussian sample
would require another argument.

\paragraph{From paths to the tests.}
For root CvM assume that the empirical probability measure of the fitted
grid, measured from the intercept, converges weakly to $Q$ on $T$. The
limiting statistic for column $j$ is $(\int G_j^2\,dQ)^{1/2}$, where $G$ is
the limit process. For KS assume that the grid converges in Hausdorff
distance to a compact $D\subseteq T$. Its limit is $\sup_{t\in D}|G_j(t)|$.
Here the limits $Q,D$ are fixed given the design. If each chosen statistic
has a continuous limiting distribution, the primary row and any fixed
number $B$ of secondary rows converge jointly to independent copies of the
same row. A single-column test therefore has limiting rejection
probability $\lfloor\alpha(B+1)\rfloor/(B+1)$. The minimum-$p$ rule over
the scales has the bounds of Proposition~\ref{prop:finite-rank-main} up to
$o(1)$. For $B_n\to\infty$ the minimum-$p$ value converges to the uniform
law by Supplementary Proposition~\ref{S-prop:joint-statistic-ranks}. These
conclusions allow perfect dependence between columns. The proof, including
the replacement of the fitted weights, is in
Section~\ref{S-sec:si-fixed-path}.

\subsection{Two balanced designs in which the conditions hold}
\label{sec:balanced}
In two designs the conditions of Theorem~\ref{thm:fixed-path} can be
checked directly, and its conclusions for the tests hold with
\emph{indicator weights as well as relative smooth scales}. In the
\emph{nested design} each subject in a centre is observed once at each of
$d\ge3$ distinct fixed values $x_1,\ldots,x_d$, for example at $d$
visits. In the \emph{crossed design} every subject meets every item at
the same $d$ values. The model has centre and subject intercepts, or
subject and item intercepts, and the fixed prediction
$\beta_0+\beta_1x$ with $\beta_1>0$. The $d$ observations of one
subject, or of one subject--item pair, form a \emph{profile}; there are
$m$ profiles.

Two facts give the result. First, the projection acts within profiles. A
weight depends on an observation only through its value of $x$, so a
weight vector repeats one $f\in\R^d$ in every profile, and $\bm R$ maps
it to the repeated copies of $P_df$, where $P_d$ projects off the
intercept and $x$ in $\R^d$. Indeed, these copies sum to zero within
every profile, where every intercept column is constant, and are
orthogonal to $x$, and the removed component is in $\col(\bm X)$. Hence
each projected coefficient is bounded, and the kernel on the grid is
$d^{-1}(P_df_s)^{\mathsf T}P_df_t$ for every $m$.

Second, the fitted ordering is the ordering by $x$. After centring $x$,
the intercept columns are orthogonal to it, so the slope estimator at any
feasible working covariance equals
$\beta_1+\sum_ix_i\epsilon_i/\sum_ix_i^2$ and has the true sign with
probability tending to one as $m\to\infty$. On that event the ranks of
the ordering values are the ranks of $x$. The fitted intercept shifts all
ordering values alike and changes no weight, and at relative scales the
slope cancels between the fitted differences and their central range.
The weights on the grid are then deterministic, so the central limit
theorem applies with fixed coefficients. The conclusions on the
fitted grid follow when the central range of the $x_k$ is positive and
the statistic has a nondegenerate limit. The residual rank $r$ is
$md-m-1$ in the nested design and $dm_1m_2-m_1-m_2$ for $m_1$ subjects
and $m_2$ items in the crossed design.
Section~\ref{S-sec:si-fixed-path} gives the corresponding proof for unit
multipliers.

Elsewhere the conditions remain assumptions.
Theorem~\ref{thm:fixed-path} imposes no convergence rate on the
intercept, which matters with a fixed number of centres, whose realised
mean effect does not average out, and its slope rate is mild: an error of
order $n^{-1/4}$, as with random slopes on crossed factors, still meets
it. For a particular ML or REML fit the slope rate and the kernel and
grid conditions remain to be verified; the argument above does not cover
the two-covariate simulations of Section~\ref{sec:sim-power}.

\paragraph{Indicator weights.}
With indicator weights, a change in slope can move observations across a
threshold, which the balanced designs exclude once the fitted slope has
the true sign. For a multivariable ordering, Section~\ref{S-sec:si-indicator-order} gives a
sufficient condition in terms of the number of observations near a
threshold, under sub-Gaussian errors. Theorem~\ref{thm:fixed-path} by
itself gives no guarantee for indicator weights.

\paragraph{What a rejection means.}
If the random-effect covariance is wrong inside the fitted span,
$\bm R\bm Z=0$ still removes the entire random contribution, and
Theorem~\ref{thm:fixed-path} continues to hold under its condition on the
fit. An omitted random coefficient with design outside that span changes
the covariance of the marks to $\bm R\bm\Omega\bm R$, where $\bm\Omega$ is
the actual response covariance. At a fixed ordering the mean path can
still be zero. A single estimate of residual variance cannot generally
reproduce the resulting kernel across thresholds.
Section~\ref{sec:unit-reference} uses an omitted curvature that is specific
to each subject--item pair, in profiles of three points, to show when a
construction can recover the variance of the path.

\section{Unit multipliers}
\label{sec:unit-reference}

Fresh draws start from an independent draw for each observation. When
repeated measurements have correlated errors, their contributions to the
path can reinforce or offset one another even when the fixed mean is
correct. To retain this dependence, we partition the observations into
\emph{units}, sum the observed contributions within each unit at every
threshold, and multiply each sum by one random draw.

The choice of unit depends on the design. The one-factor procedure of
\citet{LMMpaper} used each independent group as a unit. In the designs of
Section~\ref{sec:balanced} a natural unit is a profile. Individual
profiles can bend even when the mean trajectory is linear, and if the
errors are independent across profiles, one multiplier per profile
captures the variation that these bends contribute to the path. Coarser
units are the centres in the nested design and the subjects in the
crossed design.

For a partition into units $U$, write
$\hat c_j(t)=\bm R\bm\sigma_{\hat\lambda_j}(\hat{\bm y}^a,t)$ for the
projected weights of column $j$. The secondary process of \emph{unit
multipliers} is
\begin{equation}
 W^{*}_{j}(t)=\sum_U \xi_U\widehat\psi_{jU}(t),\qquad
 \widehat\psi_{jU}(t)=n^{-1/2}\hat c_{j,U}(t)^{\mathsf T}
                                   (\bm R\bm y)_U/\widetilde\sigma ,
 \label{eq:scores}
\end{equation}
with iid multipliers $\xi_U$ of mean zero and variance one, a wild
bootstrap in the sense of \citet{Mammen1993}. We call $\widehat\psi_{jU}$
the \emph{score} of unit $U$. The scores sum to the primary path because
$\bm R$ is idempotent. One multiplier serves all thresholds and columns of
its unit. Given the data, the covariance of the secondary process is
$\sum_U\widehat\psi_{jU}(s)\widehat\psi_{kU}(t)$ for every law of the
multipliers.

\begin{samepage}
\begin{proposition}[A criterion for unit multipliers]\label{prop:unit}
Stack a fixed finite list of thresholds and columns into vectors
$\widehat\psi_U$. Suppose the corresponding vector of the primary process
tends to $N(0,K)$,
$\sum_U\widehat\psi_U\widehat\psi_U^{\mathsf T}\to_p K$ and
$\max_U\|\widehat\psi_U\|\to_p0$. Then iid centred, variance-one unit
multipliers give the same conditional limit in probability. The rank
conclusions of Theorem~\ref{thm:fixed-path} apply to statistics on this
grid with continuous marginal limits.
\end{proposition}
\end{samepage}

In the two designs of Section~\ref{sec:balanced}, take the units to be the
centres or the subjects, respectively. Every unit consists of complete
profiles. Consequently the projected weight of a unit, padded by zeros
outside the unit, is orthogonal to \emph{all} fitted columns, and its
score in \eqref{eq:scores} equals a weighted sum of its own independent
errors, divided by $\widetilde\sigma$. If the largest unit contains $o(n)$
observations, the two conditions of Proposition~\ref{prop:unit} follow
from fourth moments. In the crossed design the subjects share items, but
their scores have lost every item intercept. This argument uses the
complete profiles. For an incomplete crossed design, or an ordering that
contains an item prediction, subject labels alone give no such identity.
Units that are independent in the raw data exist when the incidence graph
of subjects and items splits into components with independent latent
effects; a central limit theorem over such units additionally requires
many components, each with a negligible contribution.

\begin{theorem}[Profile multipliers under within-profile dependence]
\label{thm:whole-profile}
Take either design of Section~\ref{sec:balanced}, with $m\to\infty$
profiles of fixed size $d$. A profile is one subject in the nested design
or one subject--item pair in the crossed design. Allow arbitrary fitted
group intercepts. Within profile $g$, let the remaining error $u_g\in\R^d$
have mean zero and covariance $\Omega_g$. Conditional on the design the
vectors $u_g$ are independent across profiles, with uniformly bounded
fourth moments and $m^{-1}\sum_g\Omega_g\to\bar\Omega$. Use the fitted
grid of the fixed-prediction ordering, with indicator weights or positive
relative scales as in Section~\ref{sec:balanced}, with positive true slope
and positive central profile range. Let $C$ stack the rows
$(P_df_j)^{\mathsf T}$ for the chosen thresholds and scales.

Let the units be the profiles, with iid centred, unit-variance
multipliers independent of the data. Then the grid vector of the primary
process and the conditional law of the secondary grid vector have the same
Gaussian limit after removal of their common factor $1/\widetilde\sigma$.
Its covariance is $K=C\bar\Omega C^{\mathsf T}/d$. The empirical covariance
of the profile scores, without that factor, converges in probability to
$K$. For statistics with continuous limiting marginals, the KS, root CvM
and minimum-$p$ tests on the fitted grid have the calibration of
Supplementary Proposition~\ref{S-prop:joint-statistic-ranks}.
\end{theorem}

The theorem allows both unequal variances and correlations of either sign
within a profile. In the example announced in
Section~\ref{sec:fixed-order}, the model omits a curvature specific to
each subject--item pair. Gaussian draws then underestimate the variance of
the path at a single threshold by a factor of two when the two variance
components are equal, and as the number of draws grows, their nominal 5\%
test rejects a correct fixed mean with limiting probability 16.6\%.
Profile multipliers recover the variance; in a Monte Carlo check they
rejected in 5.6\% of samples, against 16.4\% for Gaussian draws
(Section~\ref{S-sec:si-fixed-path}). The proof identifies exactly why
shared fitted intercepts disappear from each profile score. A further
unmodelled effect shared across profiles can invalidate that calculation
of the covariance. Projection also removes mean departures in the fitted
span before either construction is applied.

\paragraph{Single observations and leverage.}
Putting one multiplier on each observed residual in \eqref{eq:scores}
underestimates the variance. With known $\sigma$ and fixed weights, the
expected variance of the secondary process at $t$ is
$n^{-1}\sum_i c_i(t)^2R_{ii}$, whereas the target is
$n^{-1}\sum_i c_i(t)^2$. The correction divides the $i$th residual by
$\sqrt{R_{ii}}$ when this is positive. For general units, set
$A_U=(R_{UU}^{+})^{1/2}$ and replace the residual subvector in
\eqref{eq:scores} by $A_U(\bm R\bm y)_U$; this block form is related to
the adjustment of \citet{BellMcCaffrey2002}. We then speak of
\emph{corrected} unit multipliers. With known $\sigma$ their covariance is
unbiased at deterministic weights, for every partition, under independent
errors with common variance. Section~\ref{S-sec:si-fixed-path} proves the
identity and a consistency bound on a finite grid. It covers single
observations whose residual variances $R_{ii}$ are bounded away from zero
where they are positive. In a crossed design the same bound can be checked
for proposed units. It is a condition on the projected scores, not an
independence assumption on the crossed responses.

\paragraph{Random signs.}
Random signs preserve the absolute size of every unit score; Gaussian
multipliers also randomise its size, which can affect finite-sample
rejection rates although both have the same limit
(Section~\ref{sec:profile-sim}). With $m_{\rm u}$ units and an even
statistic, the all-positive and all-negative sign vectors tie, so a test
that enumerates all sign vectors has $p\ge2^{1-m_{\rm u}}$: with five
units it cannot reject at 5\%. Finite-sample exactness needs invariance
under the selected signs; the moment conditions above yield large-sample
validity.

\section{Random-effect orderings under Gaussian errors}
\label{sec:theory}

\subsection{Exact orthogonality}
For a random-effect ordering, $S\ne\varnothing$, the term
$\bm H_S\epsilon$ in the oracle ordering shares errors with the marks
$\bm R\epsilon$. The two are uncorrelated, but uncorrelated linear
combinations of independent non-Gaussian errors can remain dependent: the
sum and the difference of two iid random signs are uncorrelated, yet one
is zero whenever the other is nonzero. Gaussian
errors make the marks independent of all orderings together, so that one
can condition on the orderings.

\begin{lemma}[Exact orthogonality]
\label{lem:orth-main}
Under \eqref{eq:model-main}, conditionally on $\mathcal D_n$ and for every
set $S$ of factors, the marks $\sigma^{-1}\bm R\bm e$ are $N(\bm0,\bm R)$
and independent of the ordering $\bm y^{O_S}=\bm X\bm\beta+\bm H_S\bm e$.
\end{lemma}

Indeed $\bm R\bm V\bm H_S^{\mathsf T}=\bm R\bm C_S=0$.
More generally the predicted random contributions depend only on the
random effects and on $(\bm I-\bm R)\epsilon$. Under Gaussian observation
errors these are independent of $\bm R\epsilon$ for any independent
random-effect law. The Gaussian results below therefore hold under
\[
 H_0:\ \ \bm y=\bm X\bm\beta+\bm v+\bm\epsilon,\quad
 \bm v\in\col(\bm Z)\text{ of any law},\quad
 \bm\epsilon\sim N(\bm0,\sigma^2\bm I)\text{ independent of }\bm v .
\]

\subsection{A common limit for oracle, primary and secondary paths}
\label{sec:gaussian-paths}

\begin{theorem}[Oracle, primary and secondary paths]\label{thm:gauss-path}
Fix finitely many columns and a compact threshold interval $T$. For
their weight vectors $w_j(t)$ define $F_{j,n}(t)=n^{-1}\sum_i w_{j,i}(t)$
and $K_{jk,n}(s,t)=n^{-1}w_j(s)^{\mathsf T}\bm R w_k(t)$, using either
all oracle or all fitted weights. Suppose both kernels converge pointwise in probability to the same
$K_{jk}$, using its continuous version when random.
Assume the Gaussian error null above, $r\to\infty$, and a feasible fit
depending only on $(\bm\Pi\bm y,\|\bm R\bm y\|)$, where
$\bm\Pi=\bm I-\bm R$. For indicator columns assume both versions of
$F_{j,n}$ converge uniformly in probability to continuous functions
on $T$. For smooth columns assume scales stay in a compact subset of
$(0,\infty)$ with probability tending to one. Oracle scales may depend
on the oracle orderings and design.

The conditional laws of the oracle path vector (given all oracle
orderings and the design), of the primary path vector (given the fitting
summaries $\bm\Pi\bm y$ and $\|\bm R\bm y\|$), and of the secondary
path vector from Gaussian draws (given the data) each approach the centred
Gaussian law with kernel $K$ in bounded-Lipschitz distance in probability
on $\ell^\infty(T)^J$. An additional sufficient condition for other
fresh draws, with any iid centred, variance-one law, is
$\max_{j,i}\sup_{t\in T}|(\bm Rw_j(t))_i|/\sqrt n\to_p0$ for the fitted
weights. Smooth columns then admit every such law; for indicator columns
assume the law is sub-Gaussian, as for random signs.
\end{theorem}

A random $K$ gives an unconditional mixture of Gaussian laws; it allows,
for example, the realised effects of five centres to influence the
limiting ordering. Under the grid conditions following
Theorem~\ref{thm:fixed-path}, with possibly random limiting grids or grid
measures and almost surely continuous conditional marginal statistic laws,
the same rank conclusions follow (Section~\ref{S-sec:si-fixed-path}). For
limits on the whole real line with estimated orderings,
Sections~\ref{S-sec:shared-conditions} and~\ref{S-sec:si-asymptotic} give
separate conditions and proofs for the oracle, primary and secondary
processes (Theorems~\ref{S-thm:oracle}, \ref{S-thm:estimated}
and~\ref{S-thm:secondary}). When neither the ordering nor the kernel
stabilises, the next result still compares the primary and the secondary
process at the current sample size.

\subsection{Gaussian draws and rotations}
\label{sec:exact-link}
Let $\bm\Pi=\bm I-\bm R$ and let $\mathcal H$ be the information generated
by $\bm\Pi\bm y$ and $\|\bm R\bm y\|$. Every feasible covariance, one of the
family of \eqref{eq:model-main}, acts on the
range of $\bm R$ as a multiple of the identity, so ML and REML fits, the
fitted orderings and the relative scales are functions of
$(\bm\Pi\bm y,\|\bm R\bm y\|)$, and under $H_0$ the direction of
$\bm R\bm y$ is uniform on the unit sphere of the range of $\bm R$ given
$\mathcal H$ (Lemma~\ref{S-lem:residual-space}). This is the classical
independence of the direction of a least-squares residual from the fit and
the residual sum of squares, here for the regression of $\bm y$ on
$[\bm X,\bm Z]$; the mixed model enters only through the ordering.

\begin{proposition}[Gaussian draws for the projected processes]
\label{prop:gaussian-O-main}
Suppose $\bm y=\bm X\bm\beta+\bm Z\bm b+\bm\epsilon$ with $\bm X$ of full
column rank, $r\ge1$, and $\bm\epsilon\sim N(\bm0,\sigma^2\bm I)$
independent of $\bm b$, whose law is arbitrary, and fit a feasible
covariance by a measurable function of $(\bm\Pi\bm y,\|\bm R\bm y\|)$, with
the GLS fixed effects at that covariance, as ML
and REML fits are when ties between equally good fits are broken by a
rule that depends only on these summaries.
\begin{enumerate}[label=(\roman*),nosep]
\item Conditionally on $\mathcal H$, the primary marks
$\bm R\bm y/\widetilde\sigma$ are $\sqrt r\,\bm U_0$ and the secondary marks
of Gaussian draws are $\bm R\bm\xi_b=d_b^\#\sqrt r\,\bm U_b$, where
$d_b^\#=\|\bm R\bm\xi_b\|/\sqrt r$, $\bm U_0,\ldots,\bm U_B$
are uniform on the unit sphere of the range of $\bm R$, and
$\bm U_0,\bm U_1,d_1^\#,\ldots,\bm U_B,d_B^\#$ are independent and
independent of $\mathcal H$.
\item Fix one process, one statistic and $J$ scale columns with
$\mathcal H$-measurable scales, including indicator weights and the
relative scales $c/I_n$. Given $\mathcal H$, the conditional laws of the
primary and a secondary process vector have bounded-Lipschitz distance
(over functions bounded by one with Lipschitz constant at most one in the
product supremum norm) at most $(24J/r)^{1/2}$.
\item Let $\widehat v_j$ be the maximum (KS) or the average (CvM) over the
fitted ordering values of the variance
$n^{-1}\bm\sigma_j(\widehat y_i^a)^{\mathsf T}\bm R\,\bm\sigma_j(\widehat y_i^a)$
of the secondary process of column $j$, with $\bm\sigma_j(t)$ the weight
vector of that column. If $r\to\infty$ and the secondary processes do not
degenerate, $(\min_j\widehat v_j)^{-1}=O_p(1)$, then for fixed $B$ the
single-column test has rejection probability
$\lfloor\alpha(B+1)\rfloor/(B+1)+o(1)$ and the minimum-$p$ rule over the
$J$ columns satisfies the bounds of Proposition~\ref{prop:finite-rank-main}
up to $o(1)$; and for $B=B_n\to\infty$ both $p$-values converge to $U(0,1)$.
\end{enumerate}
\end{proposition}

By (i), a secondary path from Gaussian draws differs from a copy of the
primary path only by the radial factor $d_b^\#$, whose square has the law
$\chi_r^2/r$. The variance in (iii) prevents the rank
comparison from degenerating, and the same argument applies to columns
from several processes. For rotations \eqref{eq:exact-draw-main} the
radial factor is one: this is the rotation test of \citet{Langsrud2005}
on the residual space, whose exactness and small-sample behaviour in
mixed models \citet{exactpaper} establishes. It is the benchmark here and
is preferable when a small $r$ makes Gaussian draws inaccurate
(Section~\ref{S-sec:si-exact}).

\subsection{Corrected unit multipliers}

\begin{proposition}[Corrected unit multipliers under Gaussian errors]\label{prop:gauss-unit}
Under the Gaussian error null and the fitting rule of
Proposition~\ref{prop:gaussian-O-main}, consider finitely many fitted
weight vectors $w_j$ given the fitting summaries $\mathcal H$. Use
the corrected unit multipliers of Section~\ref{sec:unit-reference}.
Write $c_j=\bm Rw_j$, $b_n=\max_{j,U}\|c_{j,U}\|^2$ and
$\chi_n=\|\operatorname{diag}_U(A_U)\,\bm R\,
\operatorname{diag}_U(A_U)\|_{\rm op}$. Given $\mathcal H$, the
covariance of their secondary process is unbiased for the kernel of
Gaussian draws. If $\chi_n^2 b_n/n\to_p0$ and $r\to\infty$, the
finite-dimensional law of their secondary process given the data
approaches that of Gaussian draws in bounded-Lipschitz distance in
probability, for any iid centred, variance-one multipliers.
For smooth columns in Theorem~\ref{thm:gauss-path} this extends to path
convergence on $T$ if the bound holds at every finite list of thresholds.
\end{proposition}

Here the Gaussian assumption concerns the errors of the data, not the
multipliers. It permits conditioning on a random-effect ordering. A
partition of a crossed design into subjects can be assessed by the two
quantities $b_n$ and $\chi_n$, even when the raw subject vectors share
item effects. If $\bm R$ is block diagonal over the chosen units, the
correction reduces to the identity on each residual block. Independent
random signs on those blocks then preserve the law of the residual
direction and its length, and the upper bound of
Proposition~\ref{prop:finite-rank-main} holds exactly at every sample
size, ties included (Section~\ref{S-sec:si-fixed-path}).

\section{What the tests detect}
\label{sec:detection}

A test without a named alternative should say which departures it can see.
Suppose the model of Section~\ref{sec:model} is fitted but the data are
generated by
\begin{equation}
 \bm y_n=\bm\mu_n+\bm\eta_n,\qquad
 \E{\bm\eta_n\mid\mathcal D_n}=\bm0,\qquad
 \Cov{\bm\eta_n\mid\mathcal D_n}=\bm\Omega_n,
 \label{eq:alternative-main}
\end{equation}
where $\mathcal D_n$ may contain covariates that the model omits and
$\bm\Omega_n$ is positive definite; under the model,
$\bm\mu_n=\bm X_n\bm\beta$ and $\bm\Omega_n=\sigma^2\bm V_n$. Let
$\bm\theta_n^*=(\bm\beta_n^*,\sigma_n^*,\bm D_{1,n}^*,\ldots,\bm D_{L,n}^*)$
be design-measurable pseudo-true parameters, the values that
$\widehat{\bm\beta}$, $\widetilde\sigma$ and the $\widehat{\bm D}_l$ approach
under the alternative, and mark with a star the quantities of
Section~\ref{sec:model} evaluated at them, with
$\bm e_n^*=\bm y_n-\bm X_n\bm\beta_n^*$. For $a=O_S$ the oracle process at
the pseudo-true parameters, $W_n^{a*}$, has the ordering
$\bm y_n^{a*}=\bm X_n\bm\beta_n^*+\bm H_{n,S}^*\bm e_n^*$ and the marks
$\bm e_n^{O*}=\bm R_n\bm e_n^*/\sigma_n^*$, and its \emph{mean function}
\begin{equation}
 m_n^a(t)=\frac1n\E{\bm\sigma_\lambda(\bm y_n^{a*},t)^{\mathsf T}\bm e_n^{O*}
   \bigm|\mathcal D_n}
 \label{eq:mean-function-main}
\end{equation}
is the drift of the process divided by $\sqrt n$. With indicator weights,
$m_n^a\equiv0$ says that the mark of a randomly chosen observation is mean
independent of its ordering value, given the design.

\begin{proposition}[Mean functions under Gaussian alternatives]
\label{prop:mean-main}
Assume \eqref{eq:alternative-main} with
$\bm\eta_n\mid\mathcal D_n\sim N(\bm0,\bm\Omega_n)$ and let $a=O_S$. Define
the bias and the mark--ordering covariances
\[
 \bm w_n=\frac1{\sigma_n^*}\bm R_n(\bm\mu_n-\bm X_n\bm\beta_n^*),
 \qquad
 \gamma_{n,i}^a=\Cov{e_{n,i}^{O*},y_{n,i}^{a*}\mid\mathcal D_n}
  =\frac1{\sigma_n^*}\bigl(\bm R_n\bm\Omega_n\bm H_{n,S}^{*\mathsf T}\bigr)_{ii},
\]
and let $\nu_{n,i}$ and $s_{n,i}^2$ be the conditional mean and variance of
$y_{n,i}^{a*}$. Then, for every $t$,
\begin{equation}
 m_n^a(t)=\frac1n\sum_{i=1}^n\Bigl\{
   w_{n,i}\,\E{\sigma_\lambda(y_{n,i}^{a*}-t)\mid\mathcal D_n}
   -\gamma_{n,i}^a\,\kappa_{n,i}(t)\Bigr\},
 \label{eq:mean-formula-main}
\end{equation}
where $\kappa_{n,i}(t)=-\E{\sigma_\lambda'(y_{n,i}^{a*}-t)\mid\mathcal D_n}\ge0$
for finite $\lambda$ and $\kappa_{n,i}(t)$ is the $N(\nu_{n,i},s_{n,i}^2)$
density at $t$ for $\lambda=\infty$; terms with $s_{n,i}=0$ have
$\gamma_{n,i}^a=0$. In particular:
\begin{enumerate}[label=(\alph*),nosep]
\item For $O_\varnothing$ the covariance term is absent and
$m_n^{O_\varnothing}(t)=n^{-1}\sum_iw_{n,i}\,
\sigma_\lambda((\bm X_n\bm\beta_n^*)_i-t)$, whatever $\bm\Omega_n$: the
process accumulates the projected bias along the fixed prediction.
\item Under the model, at the true parameters, $\bm w_n=\bm0$ and
$\gamma_{n,i}^a=0$ for every $i$ and every $S$, so $m_n^a\equiv0$.
\item For every $S$ and every $\bm\Omega_n$, $\sum_i\gamma_{n,i}^{O_S}=0$. A
misspecified covariance with a correctly specified mean therefore
contributes to the mean function only through differences between the laws
of the ordering values: if the mean is correct and all $y_{n,i}^{O_S*}$
have the same conditional law, $m_n^{O_S}\equiv0$.
\item If $\bm\mu_n\in\col[\bm X_n,\bm Z_n]$, then $\bm w_n=\bm0$.
\end{enumerate}
\end{proposition}

Part (a) also holds for non-Gaussian observations of finite first
moment, since the oracle weights are deterministic. A vanishing drift
alone says nothing about the level of a test of the fixed mean, because
the covariance of the marks can still be wrong
(Section~\ref{sec:unit-reference}). Part (c) explains why an omitted
random slope may be visible along a random-effect ordering. Part (d)
includes an omitted group-level covariate in a model with group
intercepts, which the projection removes together with the group
offsets.

A nonvanishing mean function gives consistency under three conditions on
the process $a$:
\begin{description}[style=sameline,leftmargin=3.2em,labelwidth=2.8em,labelsep=0.4em,font=\normalfont]
\item[(P1)] a signal: there are $c_m>0$ and design-measurable thresholds
$t_n$ with $\Pr\{|m_n^a(t_n)|\ge c_m\}\to1$;
\item[(P2)] concentration:
$n^{-1}\bm\sigma_\lambda(\bm y_n^{a*},t_n)^{\mathsf T}\bm e_n^{O*}-m_n^a(t_n)\convP0$;
\item[(P3)] plug-in: $n^{-1/2}\sup_t|\widehat W_n^a(t)-W_n^{a*}(t)|\convP0$.
\end{description}
(P3) asks the primary process to differ from the pseudo-true one by
$o_p(\sqrt n)$ only.

\begin{theorem}[Consistency]
\label{thm:consistency-main}
Let the data follow \eqref{eq:alternative-main}, let the fit exist with
probability tending to one, and assume \textup{(P1)--(P3)} for a process
$a=O_S$. Then $n^{-1/2}T_{\rm KS}(\widehat W_n^a)\ge c_m/2$ with probability
tending to one, where $T_{\rm KS}(W)=\sup_t|W(t)|$, whereas
$T_{\rm KS}(\widehat W_n^{a\#})$ is tight given the data, uniformly in $n$
and in the data, for Gaussian draws and for rotations. Consequently the
conditional probability that a secondary statistic is at least the
primary one tends to zero, and $\Pr(\widehat p\le\alpha)\to1$ for fixed
$B$ with $\alpha(B+1)\ge1$. The same holds for the grid KS statistic of
\eqref{eq:statistics-main}: with indicator weights without further
conditions, and at a finite scale if some data-measurable fitted ordering
values $g_n$ satisfy
$n^{-1/2}|\widehat W_n^a(g_n)-\widehat W_n^a(t_n)|\convP0$.

The same consistency conclusions hold for fresh draws with any iid
centred, variance-one law. For unit multipliers, let
$g_U=(Ry)_U/\widetilde\sigma$ in \eqref{eq:scores}, or
$g_U=A_U(Ry)_U/\widetilde\sigma$ for corrected unit multipliers.
A sufficient condition under the alternative is
$\max_U\|g_U\|^2\log^2(2n)=o_p(n)$ for KS. For root CvM under the
integrated signal condition below, $\max_U\|g_U\|^2=o_p(n)$ suffices.
\end{theorem}

The proof is in Section~\ref{S-sec:si-detection}; the bound on the
secondary process does not use the null hypothesis. For the root CvM
statistic the same conclusions hold when (P1) and (P2) are replaced by a
signal over a set and uniform concentration, and the minimum-$p$ rule over
$J$ columns that use the same draws is consistent when one column is and
$J\le\alpha(B+1)$ (Section~\ref{S-sec:si-detection}). With a relative
scale, $\widehat W_n^a$ is taken at the fitted scale and $W_n^{a*}$ and
$m_n^a$ at a design-measurable one, and (P3) includes the change of scale.
Under the model (P1) fails, as it must, by
Proposition~\ref{prop:mean-main}(b).

\section{Numerical results}\label{sec:simulations}
\subsection{Fresh draws and unit multipliers under skewed errors}
\label{sec:profile-sim}
We simulated the two designs of Section~\ref{sec:balanced} at $n=600$,
with five covariate values per profile and 40 units: 40 centres of three
subjects in the nested design, and 40 subjects who each meet three items
in the crossed design. The errors were independent centred exponentials.
Each sample was tested against its fitted line, using $O_\varnothing$,
root CvM and a single relative scale $c=2$.
Table~\ref{tab:fixed-reference} compares three laws for fresh draws with
two laws for unit multipliers, the units being the centres or the
subjects. Details and the underlying counts are in
Section~\ref{S-sec:si-profile-sim}.

\begin{table}[ht]
\centering
\caption{Rejections (\%) at nominal 5\% in 2000 samples per row, with
$B=199$ secondary processes per test. The mean is $x$ plus the stated
curvature times $x^2$. Monte Carlo standard errors are about 0.5
percentage points at 5\% and 1.1 at 35\%.}
\label{tab:fixed-reference}
\begin{tabular}{@{}llrrrrr@{}}
\toprule
 & &\multicolumn{3}{c}{Fresh draws}&\multicolumn{2}{c}{Unit multipliers}\\
\cmidrule(lr){3-5}\cmidrule(l){6-7}
Design & Curvature & Gaussian & Signs & Mammen & Signs & Gaussian\\
\midrule
Nested & 0 & 5.55 & 5.95 & 5.50 & 5.75 & 5.70 \\
Crossed & 0 & 4.95 & 4.65 & 4.85 & 4.00 & 3.65 \\
Nested & 0.04 & 34.80 & 34.35 & 34.70 & 32.70 & 32.35 \\
Crossed & 0.04 & 33.30 & 33.40 & 33.20 & 33.15 & 32.25 \\
\bottomrule
\end{tabular}

\end{table}

Fresh signs and draws from the asymmetric two-point law of
\citet{Mammen1993} gave power within half a percentage point of Gaussian
draws in both designs. In the crossed design, unit multipliers were more
conservative than Gaussian draws. The results support the use of several
laws for fresh draws in this example; they do not establish a general
power advantage for signs.

\subsection{Ordering and detection in fitted mixed models}\label{sec:sim-power}
A larger ML and REML study varied the fitted random effects, group counts,
orderings and scales. Its six designs include subjects within many
centres, subjects within five centres, subjects crossed with items, and a
crossed design with ten items. All tests used Gaussian draws. Across the
five main designs, root CvM with the minimum-$p$ rule rejected in
4.2--5.7\% of null samples. With an omitted quadratic term at $n=4000$,
$O_\varnothing$ rejected in 48--83\% over the six designs, compared with
6--22\% for $O$. With an omitted subject slope in the nested design, the
subject ordering $O_1$ rejected in 17.7\%, 50.3\% and 98.7\% at
$n=1000,4000,16000$. Here a random-effect ordering revealed a departure
that was weak along the fixed prediction. The complete designs, power
curves and comparisons with other tests are in
Section~\ref{S-sec:si-simulations}.

\begin{figure}[tb]
\centering
\includegraphics[width=.72\linewidth]{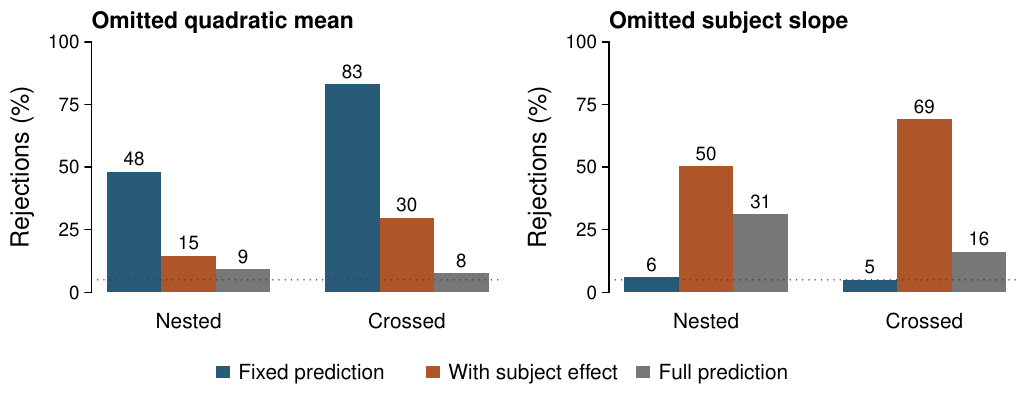}
\caption{Rejection rates at $n=4000$ for the orderings $O_\varnothing$,
$O_1$ (subjects) and $O$ under two omissions (ML, root CvM, minimum-$p$
rule over the scales, Gaussian draws). The dotted line marks the nominal
5\% level. Rates use 5000 samples, except $O_\varnothing$ under an
omitted slope, which uses 2000 samples per design. Labels give rounded
percentages.}
\label{fig:ordering-power}
\end{figure}

Figure~\ref{fig:ordering-power} holds the statistic and the construction
fixed and varies the ordering. The fixed-prediction ordering detects an
omitted quadratic mean most often in both designs. If only the subject
slope is omitted, the mean is still correct and the subject ordering gives
the highest rejection rate of the three.

\paragraph{Residual dimension and error law.}\label{sec:sim-limits}
In a constructed random-intercept and slope model with residual rank
$r=50$, Gaussian draws gave null rejection rates of 3.4--3.7\%; rotations
of the same draws gave 4.3--4.8\% (the same runs are analysed in
\citealp{exactpaper}). With independent $t_5$ or centred
$\chi^2_4$ errors in three of the designs (subjects only, subjects
crossed with items, five centres), root CvM minimum-$p$ rates with
Gaussian draws ranged from 3.9\% to 6.2\% up to $n=4000$. Under
non-Gaussian errors Theorem~\ref{thm:fixed-path} covers only the
fixed-prediction ordering with smooth weights, under its slope, design and
grid conditions; the rates for random-effect orderings and for indicator
weights in these multivariable simulations are empirical.

\FloatBarrier
\subsection{Crossed lexical-decision data}\label{sec:application}
In the \texttt{lexdec} experiment, 21 subjects responded to the same
79 words, the items of this crossed design
\citep[Section~7.1]{BaayenBook2008}. We report root CvM minimum-$p$ values
with Gaussian draws. For the logarithm of the reaction time, the model
with subject and item intercepts only was rejected along the subject
ordering and along the item ordering ($p=0.004$ and $0.010$). After
adding subject slopes for trial position and word length, the $p$-value
along the subject ordering was 0.155, along the item ordering 0.008 and
along the fixed-prediction ordering 0.006. The joint minimum-$p$ value
over the orderings of that model was 0.025; for the inverse of the
reaction time it was 0.259. The transformation and the models were chosen
on the same data.

Adding subject slopes removed the evidence of lack of fit along the
subject ordering; the item and fixed-prediction orderings still call for
examining the transformation and the mean specification. Each subject saw
each item once, so there are no repeated profiles and
Theorem~\ref{thm:whole-profile} does not apply: the rejection along the
fixed-prediction ordering cannot by itself distinguish curvature of the
mean from covariance that remains after the projection.
Section~\ref{S-sec:si-lexdec} gives the paths, the comparison with
rotations, the model specifications and the sensitivity to unequal error
variance.

\section{Discussion}\label{sec:discussion}

The ordering determines how far the multilevel construction extends beyond
Gaussian errors. With the fixed-prediction ordering the oracle path is a
linear form in independent errors. The primary path inherits its limit
when the projected coefficients are diffuse and fitting changes the
weights slowly enough once the common intercept is removed. A
random-effect ordering itself uses the observation errors. Centred
secondary marks cannot reproduce a systematic shift caused by dependence
between a mark and its position in the ordering, as can arise in small
groups with fitted random slopes.

Fresh draws for the fixed-prediction ordering assume a common error
variance and a random component that lies in the fitted span. An omitted
random coefficient outside that span changes the variance of the primary
path, which fresh draws do not follow; a mean term inside the span is
removed by the projection. For a test of the marginal mean under a potentially
wrong covariance, one must retain informative mean contrasts and estimate
their covariance under the actual dependence. For a test of the full
fitted model, the different orderings locate where the discrepancy is
visible, while its cause requires a model-based follow-up.

The constructions offer different guarantees. Under Gaussian errors, Gaussian draws are valid in
any design for large $r$, and the rotation test is exact for every $r$. When complete profiles have errors that are independent across
profiles, one multiplier per profile estimates the covariance of the
fixed-prediction path even when the covariance within a profile is
misspecified. The argument requires enough independent profiles for a
single profile to have negligible influence. In a crossed design, whether
such units exist depends on the covariate profiles and on the proposed
partition. For random slopes of crossed factors and non-Gaussian errors,
the results above give no secondary construction for a random-effect
ordering that includes those slopes.

\paragraph{Proofs and computation.}\label{sec:lean}
The proofs are formalised in Lean~4. For every numbered statement of this
paper a Lean theorem states the printed conclusion under the printed
hypotheses and is checked by the Lean kernel; Section~\ref{S-sec:si-lean}
names the theorems and describes the formalisation of the supplement.

\backmatter
\bmhead{Supplementary information}
The Supplementary Information, cited with the prefix S, contains the
proofs, the simulation protocol and the tables of the application
(Section~\ref{S-sec:si-simulations}), and the Lean formalisation of
the proofs (Section~\ref{S-sec:si-lean}).
\bmhead{Acknowledgements}
The author thanks Nata\v{s}a Kej\v{z}ar and Rok Blagus, the co-authors of
\citet{LMMpaper}, for discussions that shaped the processes studied here.

\section*{Statements and Declarations}
\bmhead{Funding}
This work was supported by the Slovenian Research and Innovation Agency
(ARIS; programme P3-0154).

\bmhead{Conflict of interest}
The author has no competing interests to declare that are relevant to the
content of this article.

\bmhead{Data availability}
The lexical decision data \texttt{lexdec} are distributed in the R package
\texttt{languageR}, version 1.6 \citep{languageR}, under the GNU General
Public License; the files read by the analysis, with checksums and source
address, are in the \texttt{simulations/} directory of the repository
named below. Simulated data
are reproducible from the recorded seeds.

\bmhead{Code availability}
The code is available at
\url{https://github.com/JakobPeterlin/multi_level_LMM_GOF}. Its
\texttt{simulations/} directory contains the
implementation of the tests (Julia package \texttt{GOFLMM} 0.1.0), the
pinned Julia 1.12 environment, and the scripts and summarised results.
The repeated-profile experiments and Figure~\ref{fig:intro-projection} use base R. The \texttt{lean/} directory contains the Lean~4 project with its
check script (Section~\ref{S-sec:si-lean}).

\bmhead{Use of large language models}
Large language models were used in the preparation of this work in 2026:
OpenAI's Astra and Sol and Anthropic's Claude Opus and Claude Fable.
They assisted in drafting and editing the text, in developing and checking
the proofs, in writing the simulation and analysis code, and in writing the
Lean formalisation of the proofs described in Section~\ref{sec:lean} and
the Supplementary Information, Section~\ref{S-sec:si-lean}; the formal
verification itself is performed by the Lean proof checker, not by a
language model, and no numerical result depends on the formalisation. The author is responsible for the content of the manuscript.

\bmhead{Author contributions}
Single author.

\bmhead{Ethics approval, consent and materials availability}
Not applicable.

\bibliography{refs}
\end{document}